\documentclass[11pt]{article} 
	\usepackage[margin=1 in, bottom=1 in]{geometry} 
	\usepackage[english]{babel} 
	\usepackage{authblk}
	\usepackage{amsmath, amssymb, textcomp, amsthm, mathtools,mathrsfs}  
	\usepackage{tensor}
	\usepackage{blkarray}
	\usepackage{wasysym} 
	\usepackage{stmaryrd}
	\usepackage{dsfont}
	\usepackage{hyperref}
   	\usepackage{natbib}
	\usepackage[utf8]{inputenc} 
	\usepackage{csquotes}
	\usepackage{footnote,footmisc} 
	\usepackage{setspace} 
	\usepackage{threeparttable}
	\usepackage{multirow}
	\usepackage{tikz}
	\usepackage{pgfplots}
	\usetikzlibrary{shapes.misc}
	\usetikzlibrary{patterns}
	\usepackage{xcolor}
	\hypersetup{
    	colorlinks,
    	linkcolor={red!50!black},
    	citecolor={blue!50!black},
    	urlcolor={blue!80!black}
	}
	\usetikzlibrary{decorations.pathreplacing}
	\usetikzlibrary{arrows}
	\usepackage{float} 
	\newcommand{\GG}[1]{}
   	 \allowdisplaybreaks
	\usepackage[toc,page]{appendix}

	\newtheorem{theorem}{Theorem}
	\newtheorem{lemma}{Lemma}

	\newtheorem*{conjecture*}{Conjecture}
	\newtheorem{definition}{Definition}

	\newtheorem*{example*}{Running Example}
	\newtheorem*{algorithm*}{Algorithm}
	\usepackage{graphicx}  
	\usepackage{marvosym} 
	\makeatletter
\def\moverlay{\mathpalette\mov@rlay}
\def\mov@rlay#1#2{\leavevmode\vtop{%
   \baselineskip\z@skip \lineskiplimit-\maxdimen
   \ialign{\hfil$\m@th#1##$\hfil\cr#2\crcr}}}
\newcommand{\charfusion}[3][\mathord]{
    #1{\ifx#1\mathop\vphantom{#2}\fi
        \mathpalette\mov@rlay{#2\cr#3}
      }
    \ifx#1\mathop\expandafter\displaylimits\fi}
\makeatother

\newcommand\independent{\protect\mathpalette{\protect\independenT}{\perp}}
\def\independenT#1#2{\mathrel{\rlap{$#1#2$}\mkern2mu{#1#2}}}

    \newtheorem*{assumptions*}{\assumptionnumber}
\providecommand{\assumptionnumber}{}


\title{Non-testability of instrument validity under continuous treatments}
\author{Florian F Gunsilius}
\affil{University of Michigan\thanks{I want to thank the editor, an associate editor, and two referees for helpful comments and references. I also thank Susanne Schennach, Toru Kitagawa, and Ya'acov Ritov as well as Marinho Bertanha, Peter Caradonna, Ken Chay, Adam McCloskey, Anna Mikusheva, Marcelo Moreira, and audiences at several universities for helpful comments. All errors are mine.}}
\date{\today}
\begin{document}
\maketitle

\begin{abstract}
This note presents a proof of the conjecture in \citet*{pearl1995testability} about testing the validity of an instrumental variable in hidden variable models. It implies that instrument validity cannot be tested in the case where the endogenous treatment is continuously distributed. This stands in contrast to the classical testability results for instrument validity when the treatment is discrete. However, imposing weak structural assumptions on the model, such as continuity between the observable variables, can re-establish theoretical testability in the continuous setting.\\

\noindent  \emph{Keywords}: Bell's inequality; continuous treatment; endogeneity; hidden variable model; instrumental variable model; instrument validity; latent variable.
\end{abstract}

\section{Introduction}
Since their introduction in Appendix B of \citet*{wright1928tariff}, instrumental variables have become the main tool for identifying causal effects in empirical settings with endogeneity arising from selection on latent variables. Applications range from randomized controlled trials with imperfect compliance to classical estimation of supply- and demand systems in economics, see \citet*{stock2003retrospectives} for a historical analysis and \citet*{imbens2015causal} for a general reference. 
\begin{figure}[h!t]
\centering
\begin{tikzpicture}
\draw node at (-2,0) {$\boxed{Z}$};
\draw node at (0,0) {$\boxed{X}$};
\draw node at (2,0) {$\boxed{Y}$};
\draw node[draw=black, circle, anchor=center] at (1,1) {$U$};
\draw[->,thick] (-1.7,0) -- (-0.3,0);
\draw[->,thick] (0.3,0) -- (1.7,0);
\draw[->,thick] (0.6,0.8) -- (0,0.3);
\draw[->,thick] (1.4,0.8) -- (2,0.3);
\end{tikzpicture}
\caption{DAG representation of the instrumental variable model with outcome $Y$, endogenous treatment $X$, instrument $Z$, and latent confounder $U$.}
\label{dagmodel}
\end{figure}

An instrument for an endogenous treatment captures most of the effect of the treatment on the outcome while being exogenous itself. A random variable hence needs to fulfill two criteria for being an instrument in a hidden variable model: it needs to (i) have an influence on the treatment and (ii) be valid, i.e.~independent of the latent terms in the model and without a direct influence on the outcome. Figure \ref{dagmodel} presents a schematic of the instrumental variable model in terms of directed acyclic graphs in the sense of \citet*{pearl1995testability}. Validity, captured by missing arrows between $Z$ and $U$ as well as $Z$ and $Y$ in Figure \ref{dagmodel}, is arguably the main requirement for a random variable to be an instrument.

The question of whether instrument validity is testable has been of interest since the introduction of instrumental variables. In this context, testability is understood in the theoretical sense of existence of restrictions on the data-generating process induced by the model. In other words, if there exist data-generating processes of the observable variables $Y$, $X$, and $Z$ which cannot be replicated by the model structure in Figure \ref{dagmodel}, then the model is theoretically testable, i.e.~falsifiable. If, on the other hand, the model is too general in the sense that it places no restrictions on the observable data-generating process, it is not testable. 

\citet*{pearl1995testability} was the first to explicitly address questions of instrument validity in a general setting. He derived an ``instrumental inequality'', a necessary theoretical condition for an instrument to be valid. This inequality requires the endogenous treatment $X$ to be discrete and led Pearl to conjecture that testability is not possible when $X$ is continuous. Since then, several results concerning the testability of instrument validity have been derived. \citet*{manski2003partial} arrives at the same instrumental inequality in the missing data context. \citet*{kitagawa2015test} derives a test when the outcome is continuous and treatment and instrument are binary, also testing monotonicity of the instrument. \citet*{wang2017falsification} derive practically useful tests of instrument validity in the binary case. \citet*{kedagni2018sharp} show the necessity and sufficiency of Pearl's conjecture in the case where all variables are binary and augment Pearl's inequality in the case where $Z$ is discrete, $Y$ is general, and $X$ is binary. \citet*{jiang2020measurement} derive sharp bounds for the binary model under the assumption of measurement error. Finally, \citet*{bonet2001instrumentality} provides a proof of Pearl's conjecture in the special case where the outcome and the instrument are discrete. 

In spite of these advancements, the question concerning the testability of the validity of instruments in general instrumental variable models has remained open in the case of continuous endogenous variables, which is a common setting in applied research: from sensitivity curves and dose-response functions in clinical trials to equilibrium models in economics. This note addresses this issue by providing a proof of Pearl's conjecture \citep*{pearl1995testability} in the most general setting. It shows that an instrumental variable model without any structural restrictions on the relations between the observable variables is too general for inducing testable implications for instrument validity when the endogenous variable is continuous. On a more positive note, we argue that weak assumptions like continuity or monotonicity between the observables re-establish theoretical testability. This provides a first na\"ive answer to the other open questions in \citet*{pearl1995testability}, asking if differentiability or monotonicity in the relation between the observable variables re-establishes theoretical testability. 

\section{Structural form of the instrumental variable model}
For the purposes of this note, it is convenient to represent the instrumental variable model depicted in Figure \ref{dagmodel} as a structural model. The structural form of the model is \citep*{pearl1995testability}
\begin{equation}\label{mainmodel}
\begin{aligned}
Y &= h(X,U)\\
X &=g(Z,U),\qquad Z\independent U
\end{aligned}
\end{equation}
where $Y$ is the outcome variable of interest, $X$ is the endogenous treatment, $Z$ is the potential instrument, and $U$ is the latent confounder. We use the terms in the singular by referring to the outcome, treatment, and instrument, even though the variables can be of arbitrary dimension. Throughout, $Z\independent U$ means that $Z$ is independent of $U$, i.e.~$P_{Z,U}(A\times B) = P_{Z}(A)P_{U}(B)$ for any set of Borel sets $A$ and $B$, where $P_{Z,U}$ denotes the joint distribution of $Z$ and $U$. 
The latent variable $U$ captures the individual heterogeneity, i.e.~all unobserved but relevant variables. The treatment $X$ is endogenous in the sense that it depends on $U$. The functions $h$ and $g$ are unknown and completely unrestricted. Potential covariates of interest can be straightforwardly included in the model by conditioning on them. We also allow for the pathological case that $X$ is continuous and independent of $Z$, in which case the relevance criterion of an instrument would be violated. \citet*{pearl1995testability} already proved that instrument validity is not testable in this special case. This also shows that the weak instrument case, i.e.~the case where $X$ and $Z$ are weakly correlated, is included in our setting and has no effect on the result. Model \eqref{mainmodel} encodes the validity of the instrument $Z$ by (i) the fact that $h$ cannot be written as a function of $Z$ (which corresponds to the missing arrow between $Z$ and $Y$ in Figure \ref{dagmodel}) and (ii) the independence restriction $Z\independent U$ (which corresponds to the missing arrow between $Z$ and $U$ in Figure \ref{dagmodel}). Full independence is required because the functions $h$ and $g$ are completely unrestricted. 

Both assumptions have interpretations in the context of a double-blind clinical trial \citep*{pearl1995testability}. Here, the utilization of placebos guarantees that the treatment assignment $Z$ does not have a direct influence on the outcome process except through the actual treatment taken ($X$). In addition, randomization of the treatment ensures independence of $Z$. When randomization is not available, which is the case in observational studies for instance, instrument validity is equivalent to the independence restriction under model \eqref{mainmodel}.

\section{Pearl's conjecture}
\subsection{Statement of the result}
This section contains the statement of Pearl's conjecture and an outline of the proof. The complete proof is relegated to the supplementary material. We prove the conjecture under the most general setting by considering a non-atomic conditional law $P_{X|Z=z}$ on Polish spaces, i.e.~complete separable metric spaces. A measure $P$ on a space $\mathcal{X}$ is non-atomic if for every Borel set $A\subset\mathcal{X}$ with $P(A)>0$ there exists a Borel set $B\subset A$ with $P(A)>P(B)>0$. This level of generality allows us to consider the setting where all variables can be infinite dimensional, which could be useful in settings where dynamic considerations play a role. 

In the following, calligraphic letters denote general sets. For instance, $\mathcal{X}_z$ denotes the support of $P_{X|Z=z}$ for fixed $z\in\mathcal{Z}$, i.e.~the smallest closed set such that $P_{X|Z=z}(\mathcal{X}_z)=1$. All supports can be of different dimensions without affecting the result. A small letter in the function $g(\cdot,\cdot)$ denotes the realization of the corresponding random variable. For instance, $g(z,U)$ denotes the map transporting the law $P_U$ to the law $P_{X|Z=z}$ for the realization $z$.

With these preparations, we can state Pearl's conjecture. 
\begin{theorem}\label{pearlsconjecture}
Let $Y$, $X$, and $Z$ be observable random variables with corresponding conditional law $P_{Y,X|Z}$. If the marginal $P_{X|Z}$ of $P_{Y,X|Z}$ is non-atomic, then there exist functions $g(Z,U)$ and $h(X,U)$ in model \eqref{mainmodel} with $U$ uniformly distributed on the unit interval, $U\independent Z$, and $g$ invertible in both $Z$ and $U$, such that the probability measure induced by this model coincides with $P_{Y,X|Z}$. 
\end{theorem}
The above statement is more technical than the wording of Pearl's original conjecture; in particular, Pearl stated that ``if [$X$] is continuous, then every joint density $f_{Y,X|Z=z}$ can be generated by the instrumental process defined in [model \eqref{mainmodel}]''. Since we work in an instrumental variable model, the relevant probability measures are $P_{Y,X|Z=z}$ and $P_{X|Z=z}$. Theorem \ref{pearlsconjecture} is also stronger than Pearl's original conjecture in that it shows that testability cannot be re-established by simply assuming $g(Z,U)$ is invertible in $U$, an assumption that is sometimes made in the literature (e.g.~\citeauthor*{dette2016testing} \citeyear{dette2016testing}, Assumption 1).

To understand why the above theorem implies that instrument validity is not testable, note that the observable distribution which gives us correct information on our causal inference problem is $P_{Y,X|Z}$. In particular, the observable $P_{Y|X}$ is biased because of the endogeneity problem between $Y$ and $X$: we are interested in the unobservable counterfactual $P_{Y(x)}$, which is the conditional distribution of $Y$ given that we fix $X=x$ exogenously. 
But if a model can produce any possible data-generating process in the form of $P_{Y,X|Z}$, then no observable distribution can induce a testable implication on the model as mentioned in the introduction.

\subsection{Outline of the proof}
To connect our proof with the conjecture in \citet*{pearl1995testability}, we call $g(Z,U)$ a generator.
\begin{definition}
Given a probability measure $P_{X|Z}$, a function $g:\mathcal{Z}\times\mathcal{U}\to\mathcal{X}\times\mathcal{Z}$ is a generator of $P_{X|Z}$ if and only if there exists some probability measure on the support $\mathcal{U}$ of $U$ such that $g(z,U)$ is distributed as $P_{X|Z=z}$ for $P_Z$-almost all $z\in\mathcal{Z}$. A generator is one-to-one if and only if $g(z_i,u)=g(z_j,u)$ implies $z_i=z_j$ for $P_Z$-almost all $z_i,z_j\in\mathcal{Z}$ and $u\in\mathcal{U}$.
\end{definition}
One-to-one generators are useful because of the following lemma, which allows us to reduce the proof of the conjecture to the first stage. This lemma is stated and proved in \citet*{pearl1995testability}, but we also provide a proof using our notation. 
\begin{lemma}\label{pearlslemma}
For any probability measure $P_{Y,X|Z}$ whose marginal $P_{X|Z}$ has a one-to-one generator $g(Z,U)$, there exists $h(X,U)$ such that $P_{Y,X|Z}$ coincides with the joint distribution produced by model \eqref{mainmodel} for these given $g(Z,U)$ and $h(X,U)$.
\end{lemma}
\begin{proof}
Let $g(Z,U)$ be a one-to-one generator of the measure $P_{X|Z}$ and factor $P_{Y,X|Z} = P_{Y|X,Z}P_{X|Z}.$ Use $g(Z,U)$ to generate $P_{X|Z}$ via $P_U$ and some other function $Y = h'(X,Z,V)$ to generate $P_{Y|X,Z}$ via some $P_{V}$, where $V\independent (U,Z)$, i.e.~$P_{U,Z,V}(A\times B\times C) = P_{U,Z}(A\times B)\cdot P_V(C)$ for all Borel sets $A$, $B$, $C$. Since $g$ is one-to-one in $z$, one can invert it to obtain $Z = g^{-1}(X,U)$ and substitute this into $h'$ to get $Y = h'(X,g^{-1}(X,U),V) = h(X,U,V)$, which conforms to model (1) if we consider $(U,V)$ as $U$. Note that $U$, $V$, and $Z$ are mutually independent since $P_{U,Z,V}(A\times B\times C) = P_{U,Z}(A\times B)\cdot P_V(C)=P_U(A)\cdot P_Z(B)\cdot P_V(C)$ by construction of $V$ and the original assumption $Z\independent U$. 
\end{proof}
Restrictions on the dimension of $U$ cannot render Lemma \ref{pearlslemma} incorrect, which follows by a classical argument about isomorphisms between Polish spaces \citep*[Theorem 9.2.2]{bogachev2007measure2}. 
Also note that by using Lemma 1 we do not make any assumptions on the distribution of $Y$, so that we can allow for general distributions of $Y$ without changing the result.

The idea of the proof of Theorem \ref{pearlsconjecture} is hence to construct for every non-atomic $P_{X|Z}$ a first stage $X=g(Z,U)$ which is invertible in $Z$. We show even more, by arguing that one can always find a $g$ that is invertible in both $Z$ and $U$. The following paragraphs describe the main idea. 

The intuition for why Pearl's conjecture is true is that for non-atomic $P_{X|Z=z}$, one can always find a continuum of different injective functions $g(z,\cdot)$ on $A_z$ and $A_u$ since every open subset contains a continuum of values. This reasoning fails in the discrete setting. We give a simple example in the next section. This is the idea for why Pearl's conjecture is true even though recent results \citep*{kedagni2018sharp} show that increasingly many restrictions are placed on the model \eqref{mainmodel} with increasing cardinality of $X$: in the continuum limit, these restrictions become vacuous.

To introduce the underlying intuition of the proof, we focus on the case where the law of $Z$, $P_Z$, is nonatomic. The formal proof in the appendix allows for general $P_Z$ with potentially countably many atoms. The challenge is to find a one-to-one generator for any possible non-atomic $P_{X|Z}$. This level of generality requires us to work in the abstract, i.e.~without specifying functional forms for $P_{X|Z=z}$ or $g$. 

The fundamental idea which helps us achieve this is to regard the function $g(z,u)$ as a continuous two-dimensional array, as depicted in Figure \ref{Condorcet_cycle_pic}. The vertical axis in Figure \ref{Condorcet_cycle_pic} indexes the continuum values of $Z$ and the horizontal axis indexes the values of $X$. Since we want our approach to work for any non-atomic $P_{X|Z=z}$, we need to work with abstract (Borel-) subsets $A_x$, $A_u$, and $A_z$ for which we assume that $x=g(z,u)$ is not one-to-one in $z$, i.e.
$g(z_i,u)=g(z_j,u)$ and is onto. In the supplement, we show that we can reduce the proof of the conjecture to the sets $A_z$, $A_u$, and $A_x$ without loss of generality. The set $A_u$ is implicitly depicted by the colour coding in Figure \ref{Condorcet_cycle_pic}, which is easiest understood vertically: the colour coding of one infinitesimal column describes for which values $z\in A_z$ the function $g(z,u)$ maps the same $u$ to the given value $x$. For instance, the left panel of Figure \ref{Condorcet_cycle_pic} is completely monochromatic on each vertical strip, meaning that for each $x\in A_x$ all $g(z,u)$ are identical for all $z\in A_z$. Different colours or shadings imply that different parts of $A_u$ are mapped to parts of $A_x$.

\begin{figure}[!htb]
\centering
\begin{tikzpicture}
\filldraw[gray!75,thick] (0,0) rectangle (0.75,1.25);
\filldraw[gray!75,thick] (0.75,0) rectangle (1.5,1.25);
\filldraw[pattern=north west lines, thick] (0,0) rectangle (0.75,1.25);
\filldraw[gray!75,thick] (0,1.25) rectangle (0.75,2.5);
\filldraw[gray!75,thick] (0.75,1.25) rectangle (1.5,2.5);
\filldraw[pattern=north west lines,thick] (0,1.25) rectangle (0.75,2.5);

\filldraw[gray!75,thick] (3.5,0) rectangle (5,1.25);
\filldraw[pattern=north west lines,thick] (3.5,0) rectangle (5,1.25);
\filldraw[gray!75,thick] (4.25,0) rectangle (5,1.25);
\filldraw[gray!75,thick] (3.5,1.25) rectangle (4.25,2.5);
\filldraw[gray!75,thick] (4.25,1.25) rectangle (5,2.5);
\filldraw[pattern=north west lines,thick] (4.25,1.25) rectangle (5,2.5);
\draw[thick] (4.25,0) -- (4.25,1.25);

\filldraw[gray!75,thick] (7,0) rectangle (7.75,1.25);
\filldraw[pattern=north west lines,thick] (7,0) rectangle (7.75,1.25);
\filldraw[gray!75,thick] (7.75,0) rectangle (8.5,1.25);
\filldraw[gray!75,thick] (7,1.25) rectangle (7.75,2.5);
\filldraw[gray!75,thick] (7.75,1.25) rectangle (8.5,2.5);

\filldraw[gray!75,thick] (10.5,0) rectangle (11.25,1.25);
\filldraw[pattern=north west lines,thick] (10.5,0) rectangle (11.25,1.25);
\filldraw[gray!75,thick] (11.25,0) rectangle (12,1.25);
\filldraw[gray!75,thick] (10.5,1.25) rectangle (11.25,2.5);
\filldraw[gray!20,thick] (11.25,1.25) rectangle (12,2.5);
\filldraw[pattern=north west lines,thick] (11.25,1.25) rectangle (12,2.5);

\filldraw[gray!20,thick] (7,1.25) rectangle (7.75,2.5);
\filldraw[gray!20,thick] (7.75,1.25) rectangle (8.5,2.5);
\filldraw[pattern=north west lines,thick] (7.75,1.25) rectangle (8.5,2.5);
\filldraw[pattern=north east lines,thick] (7,1.25) rectangle (7.375,2.5);
\filldraw[pattern=north east lines,thick] (7.75,1.25) rectangle (8.125,2.5);
\draw[very thick] (7,1.875) -- (7.75,1.875);
\draw[very thick] (7.375,1.25) -- (7.37,2.5);

\filldraw[gray!20,thick] (10.5,1.25) rectangle (10.875,1.875);
\filldraw[pattern=north east lines,thick] (10.5,1.25) rectangle (10.875,1.875);
\filldraw[gray!20,thick] (10.5,1.875) rectangle (10.875,2.5);
\filldraw[gray!20,thick] (10.875,1.25) rectangle (11.25,1.875);
\filldraw[gray!20,thick] (10.875,1.875) rectangle (11.25,2.5);
\filldraw[pattern=north east lines,thick] (10.875,1.875) rectangle (11.25,2.5);
\filldraw[pattern=north east lines,thick] (11.6255,1.875) rectangle (12,2.5);
\filldraw[pattern=north east lines,thick] (11.25,1.25) rectangle (11.625,1.875);

\draw[very thick] (0,0) rectangle (1.5,2.5);
\draw[very thick] (3.5,0) rectangle (5,2.5);
\draw[very thick] (7,0) rectangle (8.5,2.5);
\draw[very thick] (10.5,0) rectangle (12,2.5);

\draw[very thick] (0,1.25) -- (1.5,1.25);
\draw[very thick] (3.5,1.25) -- (5,1.25);
\draw[very thick] (7,1.25) -- (8.5,1.25);
\draw[very thick] (7.75,0) -- (7.75,1.25);
\draw[very thick] (8.125,1.25) -- (8.125,2.5);
\draw[very thick] (7.75,1.875) -- (8.5,1.875);
\draw[very thick] (10.5,1.25) -- (12,1.25);

\draw[very thick] (10.875,1.25) -- (10.875,2.5);
\draw[very thick] (10.5,1.875) -- (11.25,1.875);
\draw[thick] (11.25,0) -- (11.25,2.5);

\draw [decorate,decoration={brace,amplitude=7.5pt},xshift=-4pt,yshift=0pt]
(0,0) -- (0,2.5) node [black,midway,xshift=-0.7cm] 
{$A_z$};
\draw [decorate,decoration={brace,amplitude=7.5pt},xshift=-4pt,yshift=0pt]
(3.5,0) -- (3.5,1.25) node [black,midway,xshift=-0.7cm] 
{$A^2_z$};
\draw [decorate,decoration={brace,amplitude=7.5pt},xshift=-4pt,yshift=0pt]
(3.5,1.25) -- (3.5,2.5) node [black,midway,xshift=-0.7cm] 
{$A^1_z$};
\draw [decorate,decoration={brace,amplitude=7.5pt},xshift=-4pt,yshift=0pt]
(7,1.25) -- (7,1.875) node [black,midway,xshift=-0.7cm] 
{$A_z^{12}$};
\draw [decorate,decoration={brace,amplitude=7.5pt},xshift=-4pt,yshift=0pt]
(7,1.875) -- (7,2.5) node [black,midway,xshift=-0.7cm] 
{$A_z^{11}$};
\draw [decorate,decoration={brace,amplitude=7.5pt},xshift=-4pt,yshift=0pt]
(7,0) -- (7,1.25) node [black,midway,xshift=-0.7cm] 
{$A_z^{2}$};
\draw [decorate,decoration={brace,amplitude=7.5pt},xshift=-4pt,yshift=0pt]
(10.5,0) -- (10.5,1.25) node [black,midway,xshift=-0.7cm] 
{$A_z^2$};
\draw [decorate,decoration={brace,amplitude=7.5pt},xshift=-4pt,yshift=0pt]
(10.5,1.25) -- (10.5,1.875) node [black,midway,xshift=-0.7cm] 
{$A_z^{12}$};
\draw [decorate,decoration={brace,amplitude=7.5pt},xshift=-4pt,yshift=0pt]
(10.5,1.875) -- (10.5,2.5) node [black,midway,xshift=-0.7cm] 
{$A_z^{11}$};

\draw [decorate,decoration={brace,amplitude=7.5pt},xshift=0pt,yshift=-2pt]
(0,2.8) -- (1.5,2.8) node [black,midway,yshift=+.5cm,xshift=+0.15cm] 
{$A_x$};
\draw [decorate,decoration={brace,amplitude=7.5pt},xshift=0pt,yshift=-2pt]
(3.5,2.8) -- (4.25,2.8) node [black,midway,yshift=+.5cm,xshift=+0.15cm] 
{$A^1_x$};
\draw [decorate,decoration={brace,amplitude=7.5pt},xshift=0pt,yshift=-2pt]
(4.25,2.8) -- (5,2.8) node [black,midway,yshift=+.5cm,xshift=+0.15cm] 
{$A^2_x$};

\draw [decorate,decoration={brace,amplitude=7.5pt},xshift=0pt,yshift=-2pt]
(7,2.8) -- (7.375,2.8) node [black,midway,yshift=+.5cm,xshift = -0.15cm] 
{$A^{11}_x$};

\draw [decorate,decoration={brace,amplitude=7.5pt},xshift=0pt,yshift=-2pt]
(7.375,2.8) -- (7.75,2.8) node [black,midway,yshift=+.5cm,xshift = -0.05cm] 
{$A^{12}_x$};

\draw [decorate,decoration={brace,amplitude=7.5pt},xshift=0pt,yshift=-2pt]
(7.75,2.8) -- (8.125,2.8) node [black,midway,yshift=+.5cm,xshift = +0.05cm] 
{$A^{21}_x$};

\draw [decorate,decoration={brace,amplitude=7.5pt},xshift=0pt,yshift=-2pt]
(8.125,2.8) -- (8.5,2.8) node [black,midway,yshift=+.5cm,xshift = +0.15cm] 
{$A^{22}_x$};

\draw [decorate,decoration={brace,amplitude=7.5pt},xshift=0pt,yshift=-2pt]
(10.5,2.8) -- (10.875,2.8) node [black,midway,yshift=+.5cm,xshift = -0.15cm] 
{$A^{11}_x$};

\draw [decorate,decoration={brace,amplitude=7.5pt},xshift=0pt,yshift=-2pt]
(10.875,2.8) -- (11.25,2.8) node [black,midway,yshift=+.5cm,xshift = -0.05cm] 
{$A^{12}_x$};

\draw [decorate,decoration={brace,amplitude=7.5pt},xshift=0pt,yshift=-2pt]
(11.25,2.8) -- (11.625,2.8) node [black,midway,yshift=+.5cm,xshift = +0.05cm] 
{$A^{21}_x$};

\draw [decorate,decoration={brace,amplitude=7.5pt},xshift=0pt,yshift=-2pt]
(11.625,2.8) -- (12,2.8) node [black,midway,yshift=+.5cm,xshift = +0.15cm] 
{$A^{22}_x$};

\filldraw[gray!75,thick] (0,-1.5) rectangle (0.5,-1);
\filldraw[pattern=north west lines,thick] (0,-1.5) rectangle (0.5,-1);
\draw[thick] (0,-1.5) rectangle (0.5,-1) node [black,midway,xshift=+0.75cm] 
{$=A_u^1$};

\filldraw[gray!75,thick] (2,-1.5) rectangle (2.5,-1);
\draw[thick] (2,-1.5) rectangle (2.5,-1) node [black,midway,xshift=+0.75cm] 
{$=A_u^2$};

\filldraw[gray!20,thick] (4,-1.5) rectangle (4.5,-1);
\filldraw[pattern=north west lines,thick] (4,-1.5) rectangle (4.5,-1);
\filldraw[pattern=north east lines,thick] (4,-1.5) rectangle (4.5,-1);
\draw[thick] (4,-1.5) rectangle (4.5,-1) node [black,midway,xshift=+0.8cm] 
{$=A_u^{11}$};

\filldraw[gray!20,thick] (6,-1.5) rectangle (6.5,-1);
\filldraw[pattern=north west lines,thick] (6,-1.5) rectangle (6.5,-1);
\draw[thick] (6,-1.5) rectangle (6.5,-1) node [black,midway,xshift=+0.8cm] 
{$=A_u^{12}$};

\filldraw[gray!20,thick] (8,-1.5) rectangle (8.5,-1);
\filldraw[pattern=north east lines,thick] (8,-1.5) rectangle (8.5,-1);
\draw[thick] (8,-1.5) rectangle (8.5,-1) node [black,midway,xshift=+0.8cm] 
{$=A_u^{21}$};

\filldraw[gray!20,thick] (10,-1.5) rectangle (10.5,-1);
\draw[thick] (10,-1.5) rectangle (10.5,-1) node [black,midway,xshift=+0.8cm] 
{$=A_u^{22}$};

\draw[thick,->] (1.65,1.25)--(2.2,1.25);
\draw[thick,->] (5.15,1.25)--(5.7,1.25);
\draw[thick,->] (8.65,1.25)--(9.2,1.25);
\draw[thick,->] (12.15,1.25)--(12.7,1.25) node  [black,midway,xshift=+0.8cm] 
{$\ldots$};
\end{tikzpicture}
\caption{The first two steps of the iterative procedure to construct the switching map $T_zg(z,\cdot)$ on $(A_z, A_u)$.}\label{Condorcet_cycle_pic}
\end{figure}

The goal is to make this abstract $g$ invertible in $Z$ (and incidentally $U$) on all of $A_z$. Our approach is to change $g$ iteratively by a specific Cantor scheme \citep*[Definition 6.1]{kechris1995classical} for both $A_z$ and $A_u$: the idea is to split the set $A_z$ into two disjoint subsets $A_z^{1}$ and $A_z^2$ of the same size, i.e.~$P_Z(A_z^1)=P_Z(A_z^2) = \frac{1}{2}P_Z(A_z)$. This is possible because $P_Z$ is non-atomic. Since we are allowed to choose the distribution of $U$ for this proof, we assume that $P_U$ is the uniform distribution on the unit interval. We also split $A_u$ into two disjoint subsets $A_u^1$ and $A_u^2$ of the same size. This is captured in the first panel of Figure \ref{Condorcet_cycle_pic}. 

Now here is the key to the proof: since $P_{X|Z=z}$ is non-atomic by assumption, we can also set up a Cantor scheme here and split up $A_x$ into two disjoint subsets $A_x^1$, $A_x^2$ of the same size, i.e.~$P_{X|Z=z}(A_x^1)=P_{X|Z=z}(A_x^2)$ for all $z$. This is the main requirement for the construction. We now define the switching map $T^{(1)}_z:A_x\to A_x$: for $z\in A_z^2$ we let $T^{(1)}_z$ be the identity, i.e.~mapping $A_x^j$ to itself, $j\in\{1,2\}$; for $z\in A^1_z$, we define $T^{(1)}_z$ to be 
\[T_z^{(1)}g(z,A^1_{u}) = g(z,A^2_{u})\qquad\text{and}\qquad T_z^{(1)}g(z,A^2_{u}) = g(z,A^1_{u}),\] i.e.~switching $A_x^1$ and $A_x^2$. This is depicted in panel 2 of Figure \ref{Condorcet_cycle_pic}. By construction, it now holds that
\[T^{(1)}_{z_i}g(z_i,A_u^j)\neq T^{(1)}_{z_j}g(z_j,A_u^j), \quad j\in\{1,2\},\quad\text{for all $z_i\in A_z^1$ and $z_j\in A_z^2$,}\]
which is one step closer to making $T^{(1)}_zg(z,\cdot)$ the one-to-one generator.

We proceed iteratively and obtain and split the first of the subsequent subsets, i.e.~$A_z^1$ into disjoint subsets $A_z^{11}, A_z^{12}$ of the same size, as captured in panels 3 and 4 of Figure \ref{Condorcet_cycle_pic}. Defining the switching map $T^{(2)}_z$ as the identity on $A_z^{(12)}$ and as
\[T_z^{(2)}g(z,A^{\iota\wedge1}_{u}) = g(z,A^{\iota\wedge2}_{u})\qquad\text{and}\qquad T_z^{(2)}g(z,A^{\iota\wedge2}_{u}) = g(z,A^{\iota\wedge1}_{u})\] for $\iota\in\{1,2\}$, where $\iota\wedge 1$ means appending $1$ to the string $\iota$. This implies that
\[T^{(2)}_{z_i}g(z_i,A_u^k)\neq T^{(2)}_{z_j}g(z_j,A_u^k), \quad k\in\{11,12,21,22\},\quad\text{for all $z_i\in A_z^{11}$ and $z_j\in A_z^{12}$.}\]

We keep this process up iteratively, i.e.~constructing a classical Cantor scheme for $A_z, A_x$, and $A_u$, for which we have to perform this switching $T^{(n)}_z$ for every subset $A_z^{\iota}$, $\iota\in\{1,2\}^n$, $n\in\mathbb{N}$, we encounter in this process. As $n\to\infty$, this will lead the one-to-one generator as the sets $A_u^\iota$ and $A_z^\iota$ shrink to unique single points $u$ and $z$ for each $\iota\in 2^n$. The final panel in Figure \ref{Condorcet_cycle_pic} as $n\to\infty$ hence has a different colour in each infinitesimal pixel of each infinitesimal vertical strip. We can even guarantee that each infinitesimal pixel of each infinitesimal horizontal strip has a different colour, hence guaranteeing that $g$ is invertible in both $Z$ and $U$. The proof in the supplementary material contains all details.

This approach leads to a $g$ which in general does not satisfy standard regularity assumptions like smoothness or monotonicity. This is an indication that we can reinstate theoretical testability of the model if we make functional form assumptions on the $g(z,u)$, in terms of the relationship between $X$ and $Z$. We provide an argument for this in the next section.

\section{Discussion}
\subsection{Intuition for non-testability when no restrictions are placed on the model}
The verification of Pearl's conjecture is a seemingly counterintuitive result given the positive testability results derived in \citet*{pearl1995testability}, \citet*{manski2003partial}, \citet*{kitagawa2015test}, and \citet*{kedagni2018sharp}, among others. The intuition for the correctness of Pearl's conjecture lies in the complexity of the admissible models in the continuous case compared to the discrete case. \citet*{pearl1995testability} already gave an intuitive explanation for why the conjecture should be true, and we can now complement this intuition from a more rigorous perspective. 

All tests of instrument validity use the idea that if model \eqref{mainmodel} is correct and $Z\independent U$, then a change in $Z$ should not change the outcome $Y$ too drastically without changing $X$, as the latter mediates the influence of $Z$ on $Y$. As noted already by Pearl, this idea is related to Bell's inequality from quantum physics (\citeauthor{bell2004speakable} \citeyear{bell2004speakable}, \citeauthor{clauser1969proposed} \citeyear{clauser1969proposed}): in both settings the inequalities derive restrictions on the observable data-generating process $P_{Y,X|Z}$ which cannot be replicated by a model with a latent variable capturing the unobservable heterogeneity $U$ of the system. 

For instance, consider the simple setting from the remark in \citet*{pearl1995testability}, where $X$ and $Z$ each take three values, $x_1$, $x_2$, and $x_3$ as well as $z_1$, $z_2$, $z_3$, and let $U$ be uniformly distributed on the unit interval. Moreover, assume that \begin{equation}\label{positivepearl}
P_{X|Z=z_1}(x_1)+P_{X|Z=z_2}(x_1)>1.
\end{equation} In this case, the function $g(z,u)$ constructed for the first stage in the proof of Theorem \ref{pearlsconjecture} cannot exist. No matter how we partition the unit interval for $U$, the fact that $P_{X|Z=z_1}(x_1)+P_{X|Z=z_2}(x_1)>1$ always implies that there is some Borel set $A_{u}\subset[0,1]$ of probability \[P_{U}(A_{u})=P_{X|Z=z_1}(x_1)+P_{X|Z=z_2}(x_1)-1=\varepsilon_{u}>0,\] which gets mapped to $x_1$ for both $z_1$ and $z_2$, because $g(z,\cdot)$ preserves measure by construction. In the continuous setting, this example cannot hold because every Borel set of positive probability contains a continuum of points by definition, so that there is never a single point $x_1$ satisfying the above property. 

It is helpful to interpret the function $g(z,u)$ for fixed $u$ as a response profile of the unobservable unit $u$ for any given action $z$, i.e.~one path $X_z(u)\equiv g(z,u)$ of a counterfactual stochastic process $X_z$. This idea was already implicit in \citet*{pearl1995testability} and \citet*{angrist1996identification} in the binary setting. Condition \ref{positivepearl} then implies that the response $x$ is the same for the actions $z_1$ and $z_2$ for almost all units $u\in A_u$. Similarly, one has a response profile $Y_x(u)\equiv h(x,u)$ for the second stage. Together they imply a joint response profile 
\[(Y,X)_z(u)\equiv \left\{Y_{X_z(u)}(u), X_z(u)\right\}\equiv \{h(g(z,u),u), g(z,u)\},\] which takes the simple form because of the exclusion restriction. Importantly, the exclusion restriction implies that the marginal process $Y_{X_z(u)}(u)$ only depends on the position $x$ of $X_z(u)$ for given $z$ and not the whole path. The law of this joint profile $(Y,X)_z(u)$ is the one that needs to be compared to the observable data-generating-process $P_{Y,X|Z=z}$. 

Now if $g(z,u)$ is constant for some $z_1$ and $z_2$ and a given $A_u$ of positive probability as in \eqref{positivepearl}, it can happen that the observable $P_{Y,X|Z=z}$ requires a joint response profile $(Y,X)_z(u)$ in which the marginal response $Y_{X_{z}(u)}(u)$ needs to change between $z_1$ and $z_2$ for all $u\in A_u$, i.e.
\[Y_{X_{z_1}(u)}(u) \neq Y_{X_{z_2}(u)}(u).\] However, \eqref{positivepearl} implies that $X_{z_1}(u)=X_{z_1}(u)$ for these $u\in A_u$, which is a contradiction due to the fact that $Y_{X_z}$ only depends on the position $X_z$ for $z$ by the exclusion restriction. This implies immediately that the model \eqref{mainmodel} cannot replicate this specific $P_{Y,X|Z=z}$. Here, we can recognize the interplay of the exclusion restriction, which implies the form of the response profile $Y_{X_z(u)}(u)$, and the independence $Z\independent U$. At least one of the two has to be violated in the case of \eqref{positivepearl}, which implies a testable restriction on the model.

\subsection{Re-establishing testability under functional form restrictions}
The above argumentation is not possible in the setting where $P_{X|Z=z}$ is non-atomic without further assumptions. The reason is that an equation like \eqref{positivepearl} cannot hold in the continuous setting, as we condition on sets of measure zero. Hence, the observable $P_{X|Z=z}$ can never introduce a case where the response profile $X_{z}(u)$ is constant over some set $A_z\subset\mathcal{Z}$ and for some $u\in A_u$. This is the intuition of the proof of Pearl's conjecture as captured in Figure \ref{Condorcet_cycle_pic}. The observable data-generating-process $P_{X|Z=z}$ could still be very erratic in principle, but the proof of Pearl's conjecture shows that we can always find an erratic enough response profile $X_z(u)\equiv g(z,u)$ which allows us to replicate it. However, for each $u$ the constructed paths $X_z(u)\equiv g(z,u)$ in the proof of Pearl's conjecture do not satisfy regularity properties like continuity or monotonicity in general.

Therefore, one can re-establish testability of model \eqref{mainmodel} even in the continuous case if one is willing to make weak structural assumptions on the paths $X_z(u)$. In particular, by the above reasoning one can immediately obtain testable implications if one assumes that $X_z(u)$ is constant on some set $A_z\in\mathcal{Z}$ of positive probability for some set $A_u$ as above. This assumption is rather artificial. However, using the same reasoning, one can straightforwardly show that a continuity assumption on $X_z$ and $Y_x$, based on the Kolmogorov-Chentsov theorem \citep*[corollary 14.9]{kallenberg2006foundations} for instance, is already enough to introduce testable implications in principle. Intuitively, the joint response profile $(Y,X)_z(u)$ must be continuous if both $Y_x(u)$ and $X_z(u)$ are continuous, and there exist data-generating processes $P_{Y,X|Z=z}$ which would induce a response profile with jumps. A similar reasoning can be made rigorous for monotone response profiles. This gives an informal positive answer to the other open questions in \citet*{pearl1995testability} about re-establishing testability through structural assumptions like differentiability and monotonicity. More work needs to be done to analyze stronger functional form assumptions on $g$ and $h$ which make the model testable in practice, not just theoretically. 

Lastly, this article may also have implications for local hidden variable theories \citep*{bell2004speakable}. It shows that even though a Bell-type inequality does not exist in our continuous setting \citep{ou1992realization}, one has to allow for general models without functional form restrictions in order to arrive at this conclusion. 
\newpage
\appendix

\section{Proof of Theorem \ref{pearlsconjecture}}
\subsection{Set-up}
The idea of the proof will be to make Figure \ref{Condorcet_cycle_pic} formal. For this, we construct a special Cantor scheme \citep*[chapter 6]{kechris1995classical} on both $A_z$ and $A_u$ in the case where $P_Z$ is continuous, and a Lusin scheme \citep*[chapter 7]{kechris1995classical} in the case where $P_Z$ has atoms. We present these two concepts in the next section. The largest part of the proof will be to analyze the limit of these schemes, to make sure that the limit of Figure \ref{Condorcet_cycle_pic} behaves as we expect. 

In the following, we say $P$ is absolutely continuous if it possesses a density $f$ with respect to Lebesgue measure, which we denote by $\lambda$. For the proof, we use measure-preserving isomorphisms and disintegrations. The following two paragraphs contain a brief overview of these two concepts. For a formal treatment we refer to chapters 9 and 10 in \citet*{bogachev2007measure2}. 

The concept of disintegrations gives meaning to the restriction of a joint probability measure $P_{Y,X}$ to a subset of Lebesgue measure zero, for instance the conditional measure $P_{Y|X=x}$ when $X$ is a continuous random variable inducing an absolutely continuous probability measure $P_X$ with respect to Lebesgue measure. A disintegration $P_{Y|X=x}(A)$ for some Borel set $A\in\mathcal{B}_Y$ is a version of the standard conditional expectation $E(\mathds{1}\{Y\in A\}|\mathcal{F}_X)$ for some filtration $\mathcal{F}_X\subset\mathcal{B}_X$ when it exists, where $\mathds{1}\{A\}$ denotes the standard indicator function which is $1$ if the event $A$ happens and $0$ otherwise, and where $\mathcal{B}_X$ denotes the Borel $\sigma$-algebra on $\mathcal{X}$. The existence of a disintegration can be shown under very general circumstances and is guaranteed in our setting of probability measures on Polish spaces equipped with the Borel $\sigma$-algebra (see Theorem 1 in \citeauthor*{chang1997conditioning} \citeyear{chang1997conditioning}).

The second formal concept required for some of the proofs is that of a measure-preserving isomorphism \citep*[Definition 2.1]{einsiedler2013ergodic}. A map $T:\mathcal{X}\to\mathcal{Y}$ transporting a probability measure $P_X$ onto another probability measure $P_Y$ is measure-preserving if it is (i) measurable and (ii) $P_Y(A)=P_X(T^{-1}A)$ for every set $A$ in the Borel $\sigma$-algebra $\mathcal{B}_{Y}$ corresponding to $Y$. Measurability of $T$ means that $T^{-1}A\in\mathcal{B}_{X}$ for all $A\in\mathcal{B}_Y$, where $T^{-1}A$ denotes the set of points $x\in \mathcal{X}$ such that $Tx\in A$. 
If $T$ is invertible and its inverse is also measure-preserving, it is called a measure-preserving isomorphism. For all our work, we only need measure-preserving isomorphisms up to sets of Lebesgue measure zero, so that from now on we mean ``measure-preserving isomorphism modulo sets of measure zero'' when we write ``measure-preserving isomorphism''. This implies that statements like $T_zg(z, A)=T_zg(z,B)$ for Borel-sets $A$ and $B$ are implicitly understood to only hold up to sets of measure zero.

\subsection{Proof of Pearl's conjecture}
\subsubsection{Cantor- and Lusin schemes}
The main tool for the proof is to create a Cantor scheme with shrinking diameter on Borel sets, i.e.~a structured system of subsets $\{A^\iota\}_{\iota\in2^{<\mathbb{N}}}$ of some Borel set $A$ which satisfy \citep*[Definition 6.1]{kechris1995classical} 
\begin{align*}
(i)&\mbox{}\quad A^{\iota\wedge 1}\cap A^{\iota\wedge 2} = \emptyset\quad\text{for any $\iota\in 2^{<\mathbb{N}}$,}\\
(ii)&\mbox{}\quad A^{\iota\wedge i}\subset A^{\iota} \quad\text{for $i\in\{1,2\}$, and}\\
(iii)&\mbox{}\quad \lim_{n\to\infty}\text{diam}(A^{\tau\vert n}) = 0\quad\text{for $\tau\in 2^{\mathbb{N}}$},
\end{align*}
where $2^{<\mathbb{N}}\coloneqq\bigcup_{n\in\mathbb{N}} \{1,2\}^n$ denotes the set of all finite sequences of the values $\{1,2\}$, $2^{\mathbb{N}}$ denotes the corresponding Cantor space, and $\iota\wedge i$ means appending the value $i$ to the sequence $\iota\in2^{<\mathbb{N}}$. The symbol $\tau\vert n$ denotes the initial $n$-segment of an element $\tau\in 2^{\mathbb{N}}$, which is the value $\iota\in 2^{n}$ which has the same first $n$ values as $\tau$. For instance, if $\tau=121212\ldots$, then $\tau\vert 3$ denotes the sequence $121$. The diameter $\text{diam}(A)$ of a set $A$ is defined as 
\[\text{diam}(A) = \sup\{d(x,y): x,y\in A\},\] where $d$ is some metric. In our case, it will be the Euclidean distance, because we show below in Lemma \ref{isolemma} that we can restrict the the problem to the unit interval. We create such a Cantor scheme for $A_z$, $A_u$ and $A_x$ in the case where $P_Z$ is non-atomic. 
If $P_Z$ contains (possibly infinitely many) atoms, we extend the Cantor scheme to a Lusin scheme on $\mathbb{N}^{<\mathbb{N}}$ \citep*[Definition 7.5]{kechris1995classical}, where $\mathbb{N}^{<\mathbb{N}}$ is the set of all finite sequences of values in $\mathbb{N}$. The three requirements on the sets stay the same.

The main idea in our construction is that we split Borel sets $A_z^{\iota}$ and $A_u^\iota$ for $\iota\in 2^n$ into Borel sets $A_z^{\iota\wedge 1}$ and $A_z^{\iota\wedge 2}$ as well as $A_u^{\iota\wedge 1}$ and $A_u^{\iota\wedge 2}$ of the same measure at every stage $n$ in this scheme and use a measure-preserving map $T$ that switches sets in the sense of Figure \ref{Condorcet_cycle_pic}. The main part of the proof will be to prove that these maps are measurable and that throwing away a set of measure zero in each iteration does not accumulate to a set of positive measure.

\subsubsection{Lemmas}
We split the proof into several lemmas and the main proof.

The first lemma is a well-known result about the equivalence of non-atomic measures on Polish spaces and the unit interval equipped with Lebesgue measure. This allows us to reduce the proof of the conjecture to the unit interval with absolutely continuous measures. We repeat the statement here, for a proof we refer to \citet*[Theorem 9.2.2]{bogachev2007measure2}.
\begin{lemma}\label{isolemma}
Let $(\mathcal{X},\mathcal{B}, P)$ be a Polish space, i.e.~a complete separable metric space, equipped with the Borel $\sigma$-algebra and a probability measure $P$. Then there exists a measure preserving isomorphism from $(\mathcal{X}, \mathcal{B}, P)$ to $([0,1], \mathcal{B}_{[0,1]}, \mu)$, where $\mu$ is some Borel probability measure. If $P$ is non-atomic then one can take $\mu$ to be Lebesgue measure.
\end{lemma}

Our construction of $g$ in the proof of Pearl's conjecture also relies on being able to split a Borel set $A$ of cardinality continuum and size $P(A)=\varepsilon>0$ into $k$ subsets $A_1, A_2,\ldots, A_k\subset A$ of equal measure, i.e.~$P(A_1)=P(A_2)=\ldots=P(A_k)=\frac{1}{k}\varepsilon$. The fact that this is always possible is well-known. A general proof of this is can be found in \citet*[Corollary 1.12.10]{bogachev2007measure1} for instance. We want these subsets to be an ``almost partition'' in the sense that we decompose $A$ into two closed subsets $A^1, A^2$, and a residual set $N$, for which we have not found a proof in the literature. We proof this fact now, with will also give a proof of the classical decomposition result.
\begin{lemma}\label{splitlemma}
Let $([0,1],\mathcal{B}_{[0,1]},P)$ be the probability space of the unit interval equipped with the Borel $\sigma$-algebra and $P$ an absolutely continuous measure. Then one can split any Borel set $A\in\mathcal{B}_{[0,1]}$ with $P(A)>0$ into $k$ disjoint Borel sets $A_1,A_2,\ldots, A_k\subset A$ with $P(A_1)=P(A_2)=\ldots=P(A_k)=\frac{1}{k}P(A)$ for $k\geq 2$. Furthermore, one can also obtain an ``almost partition'' of $A$: for every $\varepsilon>0$ with $\varepsilon< P(A)$, there exist closed disjoint $A^1, A^2, \ldots A^k$, and some $N$ with $P(N)=\varepsilon$ such that $P(A^1) = \ldots = P(A^k) = \frac{1}{k} (P(A)-\varepsilon)$ and $\bigcup_{j=1}^k A^k\cup N=A$.
\end{lemma}
\begin{proof}
We first prove the classical result for Lebesgue measure and then for a general absolutely continuous measure. Afterwards, we prove the existence of the ``almost partition''.

In the setting of Lebesgue measure, we first focus on $k=2$, and show the more general case at the end of this paragraph. Consider the function $f(u)\coloneqq \lambda(A \cap(0,u])$. $f$ is continuous since $A$ is a Borel set of positive probability and $\lambda$ is Lebesgue measure. In fact, it holds that \[|f(u_1)-f(u_2)|\leq|u_1-u_2|\quad\text{for $u_1,u_2\in[0,1]$}.\] Furthermore, $\lim_{u\to0}f(u) = 0$ and $\lim_{u\to1}f(u) = \lambda(A)$, so that $f$ takes on any value between $0$ and $\lambda(A)$. Therefore, we can find $u^*\in[0,1]$ such that $f(u^*)=\frac{1}{2}\lambda(A)$, so that the sets $A\cap(0,u^*]$ and $A\cap(u^*,1]$ satisfy our requirement. We can use the same idea to find values $u^{(j)}\in[0,1]$, $j=1\ldots, k$, such that $f(u^{(j)})=\frac{j}{k}\lambda(A)$. The sets $A\cap(0,u^{(1)}]$, $A\cap(u^{(j)},u^{(j+1)}]$ then satisfy our requirement.

An analogous argument can be constructed when $P$ is non-atomic by using the standard quantile transformation $T(x) = F^{-1}(F_\lambda(x)) = F^{-1}(x)$ between $P$ and Lebesgue measure $\lambda$ as the measure-preserving isomorphism. $F:[0,1]\to[0,1]$ is the CDF corresponding to $P$ and $F_\lambda$ is the CDF corresponding to Lebesgue measure. The quantile function $F^{-1}$ is defined in the usual way as
\[F^{-1}(t)\coloneqq \inf\{x\in[0,1]: F(x)=t\}.\]
Since $P$ is non-atomic its associated distribution function $F:[0,1]\to[0,1]$ is continuous and also strictly increasing on its support $\mathcal{S}\subset[0,1]$ defined as the smallest set such that $P(\mathcal{S})=1$ \citep*{embrechts2013note}. Note that $F$ is only continuous on all of $[0,1]$ but not strictly increasing. We therefore need to restrict the argument to the support $\mathcal{S}$. Since $F$ is strictly increasing and continuous on its support, $F^{-1}: [0,1]\to\mathcal{S}$ is strictly increasing and continuous, so that $F$ is a measure-preserving homeomorphism (i.e.~continuous with continuous inverse) on its support $\mathcal{S}$. Again, this result is not true if we were to consider $[0,1]$ instead of $\mathcal{S}$. 

Therefore, and since $P(A)>0$, $F^{-1}$ is a measure-preserving homeomorphism on $A\cap\mathcal{S}$, so that the function $f(F(x))\coloneqq P(F(A\cap\mathcal{S}\cap(0,x]))$ is continuous which implies the existence of an $x^*$ with $f(x^*)=\frac{1}{2}P(A\cap\mathcal{S})=\frac{1}{2}P(A)$ by the same argument as the one for Lebesgue measure. The same idea as in the setting of Lebesgue measure now works for $k>2$ sets.

Finally, we argue that in fact we can consider $A^1, A^2$  to form an ``almost partition'' of $A$ in the sense that for every $\varepsilon>0$, there exist compact Borel sets $\tilde{A}^j\subset A^j$, $j\in\{1,2\}$, and a set $N$ with $P(N)=\varepsilon$ such that 
\[\tilde{A}^1\cap \tilde{A}^2 \cap N=\emptyset,\quad \tilde{A}^1\cup \tilde{A}^2\cup N= A,\quad P(\tilde{A}^1)=P(\tilde{A}^2)=\frac{1}{2} (P(A)-\varepsilon).\]
This follows from the fact that an absolutely continuous probability measure on the unit interval is a Radon measure \citep*[Theorem 7.1.7]{bogachev2007measure2}, which means that for every Borel set $A$ and every $\varepsilon>0$ there exists a closed (and hence compact) Borel set $B\subset A$ such that
\[P(A\setminus B)\leq\varepsilon.\] To use this, fix some $\varepsilon>0$ and partition $A$ into disjoint $A^1$ and $A^2$ with $P(A^1)=P(A^2)=\frac{1}{2}P(A)$ as above. Then by the fact that $P$ is a Radon measure, it holds that there exist compact $B^1\subset A^1$ and $B^2\subset A^2$ with $P(A^j\setminus B^j)\leq\frac{1}{2}\varepsilon$. In particular, $B^1$ and $B^2$ are disjoint. Note that we can assume $P(B^1)=P(B^2)$ because if $P(B^1)-P(B^2)=\delta$ for some $\delta>0$, we can approximate both $B^1$ and $B^2$ by sets $\tilde{B}^1$ and $\tilde{B}^2$ with $P(\tilde{B}^1)=P(\tilde{B}^2)$ and $P(B^1\setminus\tilde{B}^1)+P(B^2\setminus\tilde{B}^2) = \delta$ and letting $\delta\to0$. The sets $\tilde{B}^1$ and $\tilde{B}^2$ are the required sets. 
The same argument can be straightforwardly extended to the case $k>2$.
\end{proof}

Another important part of our proof of Pearl's conjecture, as depicted in Figure \ref{Condorcet_cycle_pic}, is the idea of ``switching'' sets of equal probability in the Cantor scheme. The challenge for this is to always find a measure preserving isomorphism between two Borel sets $A,B\subset[0,1]$. The following lemma, which is analogous to the Lemma on page 74 in \citet*{halmos1956lectures}, shows that this is always possible. The proof is verbatim the one on page 74 in \citet*{halmos1956lectures}, so that we omit it.
\begin{lemma}\label{halmoslemma}
Consider a probability space $([0,1], \mathcal{B}_{[0,1]}, P)$ and Borel sets $A,B\in\mathcal{B}_{[0,1]}$ with $P(A)=P(B)$. Then there exists a measure-preserving isomorphism $T: [0,1]\to [0,1]$ which maps $A$ bijectively into $B$ such that 
$P((TA\setminus B)\cup (B\setminus TA))=0.$
\end{lemma}

For any $k\in\mathbb{N}$ we will need to construct a ``cyclic map'', i.e.~for mutually disjoint Borel sets $A_1,\ldots A_k$, $k\geq 2$, we want to construct a measure-preserving isomorphism $T$ which maps 
\[TA_{j} = A_{j+1},\quad \text{for}\quad j=1,\ldots, k\quad\text{with} \quad TA_{k}=A_1.\] 
The general existence of such maps is a classical result in ergodic theory which can be shown via periodic maps \citep[p.~70]{halmos1956lectures}. However, we have not found a proof for a priori specified Borel sets $A_1,\ldots, A_k$, so that we provide a proof here.

\begin{lemma}\label{switchlemma}
Consider a probability space $([0,1], \mathcal{B}_{[0,1]}, P)$ and mutually disjoint Borel sets $A_1,\ldots, A_k\in[0,1]$ with $P(A_1)=\ldots= P(A_k)>0$, $k\in\mathbb{N}$. Then there exists a measure-preserving isomorphism $T:[0,1]\to[0,1]$ which maps 
\[TA_{j} = A_{j+1},\qquad\text{for}\quad j=1,\ldots, k\quad\text{ with}\quad TA_k=A_1\] and is the identity on $[0,1]\setminus\bigcup_{j=1}^k A_j$. 
\end{lemma}
\begin{proof}
Consider the case $k=2$ with $A_1$ and $A_2$ first. By Lemma \ref{halmoslemma} there exists a measure-preserving isomorphism $T^{(1)}:[0,1]\to[0,1]$ such that $T^{(1)}A_1=A_2$. We now restrict $T^{(1)}$ to $A_1$, i.e.~we consider
\[\tilde{T}^{(1)} = \left.T^{(1)}\right\rvert_{A_1}.\] Again by Lemma \ref{halmoslemma}, there exists a measure-preserving isomorphism $T^{(2)}:[0,1]\to[0,1]$ such that $T^{(2)}A_2=A_1$. In fact, if we only have two sets $A_1$ and $A_2$, we can define $T^{(2)}(x) = (T^{(1)})^{-1}(x)$. 
Again, define 
\[\tilde{T}^{(2)} = \left. T^{(2)}\right\rvert_{A_2}.\] The map $T:[0,1]\to[0,1]$ 
\[T =  \tilde{T}^{(1)}\cup \tilde{T}^{(2)}\cup Id\coloneqq\begin{cases}\tilde{T}^{(1)}&\text{on $A_1$}\\ \tilde{T}^{(2)}&\text{on $A_2$}\\ Id&\text{on $[0,1]\setminus (A_1\cup A_2)$}\end{cases}\] is the required map, where $Id(x) = x$ is the identity operator.

In the setting $k\geq 3$, we need to allude to Lemma \ref{halmoslemma} $k$ times to obtain maps $T^{(j)}:A_j\to A_{j+1}$ with $T^{(k)}A_k=A_1$. We then again restrict
\[\tilde{T}^{(j)} = \left. T^{(j)} \right\rvert_{A_j},\qquad\text{for all $j=1,\ldots,k$}.\] Then
\[T\coloneqq \bigcup_{j=1}^k\tilde{T}^{(j)}\cup Id\] is the required map.
\end{proof}

We are now ready to prove the conjecture.

\subsubsection{Proof of Pearl's conjecture}
\begin{proof}[Proof of Theorem 1]
We proceed in three parts by splitting the argument into the case where $P_Z$ is non-atomic (part 1), has finitely many atoms and an absolutely continuous part (part 2), and has countably many atoms and an absolutely continuous part (part 3).  All important results are contained in part 1, which is split into three further parts: part 1.1 introduces the Cantor scheme on some fixed Borel sets $A_u$, $A_z$, and $A_x$; part 1.2 provides a proof that the map constructed by the Cantor scheme is measurable and well defined; part 1.3 shows that the restriction to Borel sets is without loss of generality. In part 1 we can use a Cantor scheme with vanishing diameter on $2^{<\mathbb{N}}$ because we only need to split each subset $A^{\iota}_u$ into two further subsets $A_u^{\iota\wedge 1}$, $A_u^{\iota\wedge 2}$. The second part shows how we need to adjust our argument if $P_Z$ possesses $k<+\infty$ atoms. The key will be to define a cyclic map. In the third part, we argue that the case for countably many atoms follows from a simple approximation argument, using a general Lusin scheme on $\mathbb{N}^{<\mathbb{N}}$. 

In light of Lemma \ref{pearlslemma}, we only have to construct a one-to-one generator $g(Z,U)$ between some distribution $P_U$, which we are allowed to choose, and any possible non-atomic distribution $P_{X|Z=z}$ for $P_Z$-almost every $z\in\mathcal{Z}$. In the following, we drop the phrase ``$P_Z$-almost every'' for readability purposes whenever possible. Throughout, we will fix $P_U$ to be the uniform distribution on the unit interval $[0,1]$, independent of the distribution $P_Z$. This assumption on the distribution of $U$ will be general enough to prove Pearl's conjecture. In addition, we can assume that all random variables have an absolutely continuous distribution and take values in the unit interval, i.e.~that the support of $P_{X|Z=z}$ is in $[0,1]$ for $z\in[0,1]$. This follows directly from Lemma \ref{isolemma}.

Also notice that $g(\cdot,\cdot)$ in model (1) is by definition a family of measure-preserving maps $g(z,\cdot): [0,1]\to[0,1]$ transporting the measure $P_U$ onto the measures $P_{X|Z=z}$. Moreover, since $P_{U}$ and $P_{X|Z=z}$ are probability measures, we can even assume that these functions $g$ are measure-preserving isomorphisms, i.e.~invertible in $U$ with measure-preserving inverses $g^{-1}(X,z)$, again by Lemma \ref{isolemma}. This implies that our approach will provide a $g$ that is invertible in both $Z$ and $U$ if we make it invertible in $Z$, as the ``switching map'' $T_z$ from Lemma \ref{switchlemma} will not affect the injectivity of $g(z,\cdot)$ since it is always a measure-preserving isomorphism and hence invertible itself. \\

\noindent\emph{Part 1: Non-atomic $P_Z$}\\
We now show that a required $g$ exists in the case where $P_Z$ is non-atomic. We split this part into three further parts. In part 1.1 we construct the one-to-one generator on fixed Borel subsets $A_z$, $A_x$, $A_u$ where $g$ is not one-to-one in $Z$. In part 1.2 we prove measurability and existence of our procedure. In part 1.3 we show that it is without loss of generality to focus on fixed subsets $A_x$, $A_u$, $A_z$. \\

\noindent\emph{Part 1.1: Constructing an injective generator on fixed sets $A_z$ and $A_u$}\\
We construct $g$ by using a family of measure-preserving isomorphisms $T_z$ on the probability spaces $([0,1],\mathcal{B}_{[0,1]},P_{X|Z=z})$ for $P_Z$-almost all $z\in\mathcal{Z}$. The family $T_z$ is mapping to the same measure space, whereas $g(z,\cdot)$ maps between different measure spaces for fixed $z$. We need this set-up because we want to use Lemma \ref{switchlemma} which is based on Lemma \ref{halmoslemma} which in turn requires $T_z$ to be measure-preserving with respect to the same measure space. In order to assess whether the family $g(z,U)$, $z\in\mathcal{Z}$, forms a one-to-one generator, we have to take into account the measure on $Z$. For this we use the fact that $P_{X|Z=z}$ is a disintegrated measure, that is
\[P_{X,Z}(A_x\times A_z) = \int_{A_z} P_{X|Z=z}(A_x) P_Z(dz)\quad\text{for all Borel sets $A_x \times A_z\in\mathcal{B}_X\otimes\mathcal{B}_Z$,}\]
from which it follows that
\begin{equation}\label{disinteq}
P_{X,Z}(A_x\times A_z) = \int_{A_z} P_{X|Z=z}(A_x) P_Z(dz) = \int_{A_z} P_{U}(g^{-1}(A_x,z)) P_Z(dz)
\end{equation}
for all $A_x\times A_z\in\mathcal{B}_X\otimes\mathcal{B}_Z$. Here, $\mathcal{B}_X\otimes\mathcal{B}_Z$ is the smallest Borel $\sigma$-algebra induced by the product of the sets $A_x\times A_z$ which coincides with the product of $\mathcal{B}_X$ and $\mathcal{B}_Z$, because all sets are separable.
So in the following we will work with the map $g:(z,u)\mapsto (x,z)$
\[g: ([0,1]^2, \mathcal{B}_{[0,1]}\otimes  \mathcal{B}_{[0,1]}, P_Z\otimes P_U)\to([0,1]^2, \mathcal{B}_{[0,1]}\otimes  \mathcal{B}_{[0,1]}, P_{X,Z}),\] where $P_Z\otimes P_U$ denotes the independence coupling of $P_Z$ and $P_U$, i.e.~$P_Z\otimes P_U(A_z\times A_u)=P_Z(A_z)\cdot P_U(A_u)$ for all Borel sets $A_z$ and $A_u$. 

If $g(z,u)$ already turns out to be a one-to-one generator, there is nothing to prove and we can let $T_z$ be the identity for all $z$. So assume that $g$ is not a one-to-one generator, meaning that there are Borel sets $A_{u}\in\mathcal{B}_{[0,1]}$ and $A_z\in \mathcal{B}_{[0,1]}$ of measure $P_{U}(A_{u})=\varepsilon_{u}$ and $P_Z(A_z)=\varepsilon_z$ for some $\varepsilon_{u},\varepsilon_z>0$ such that $g(z_i,u) = g(z_j,u)$ for $u\in A_{u}$ and $z_i,z_j\in A_z$.  Since $g(z,\cdot)$ is a measure-preserving isomorphism, this is equivalent to the existence of Borel sets $A_x\in \mathcal{B}_{[0,1]}$ and $A_z\in \mathcal{B}_{[0,1]}$ such that $g^{-1}(A_x,z_i)= g^{-1}(A_x,z_j)$ for $z_i\neq z_j$. In other words, the functions $g(z,\cdot)$ indexed by $z\in A_z$ are (almost everywhere) identical on the set $A_u$ in the sense that $A_u=g^{-1}(A_x,z)$ for $z\in A_z$. In this case we can write \eqref{disinteq} as
\begin{equation}\label{disinteq2}
P_{X,Z}(A_x\times A_z) = \int_{A_z} P_{X|Z=z}(A_x) P_Z(dz) = \int_{A_z} P_{U}(g^{-1}(A_x,z)) P_Z(dz) = P_U(A_u)P_Z(A_z)
\end{equation}
due to $Z\independent U$.

We now construct the family of measure-preserving isomorphisms $T_z: A_x\to A_x$ for $A_x=g(z,A_u)$ and $P_Z$-almost all $z$, which turns $g$ into a one-to-one generator in the sense that $T_{z_i}(g(z_i, A_u)) \neq T_{z_j}(g(z_j, A_u))$ for $P_Z$-almost every $z_i,z_j\in A_z$. We construct these maps $T_z\circ g(z,\cdot)$ on $A_z$ via the iterative procedure depicted in Figure \ref{Condorcet_cycle_pic}. This is a specific Cantor scheme on $2^{<\mathbb{N}}$.

To start, partition $A_z=A_z^1\cup A_z^2$ into two closed disjoint subsets $A^1_z$ and $A^2_z$ of equal measure, i.e.
\[P_Z(A^1_z) = P_Z(A^2_z)=\frac{1}{2}(\varepsilon_Z-\delta)\] for some $\delta>0$, which we will make go to zero in part 1.2. This is always possible by Lemma \ref{splitlemma}. In particular, note that by Lemma \ref{splitlemma} $A_z^1, A_z^2$ and some residual sets $N_z$ with $P_Z(N_z)=\delta$ are mutually disjoint, where we can assume that $A_z^1$ and $A_z^2$ are closed. We also split $A_{u}$ into two closed disjoint subsets and a residual set $N_u$ of probability $P_U(N_u)=\delta'>0$ such that $A_{u}=A^1_{u}\cup A_{u}^2\cup N_u$ with 
\[P_{U}(A_{u}^1)=P_{U}(A_{u}^2)=\frac{1}{2}(\varepsilon_{u}-\delta').\] 
This is important as the switching map defined in Lemma \ref{switchlemma} requires disjoint sets. In part 1.2, we will take both $\delta,\delta'\to0$. The idea for introducing $\delta$ and $\delta'$ is very similar to proving the existence of fat Cantor sets, i.e.~Cantor sets of positive probability. We show that as we take out smaller and smaller residual sets, i.e.~taking $\delta,\delta'\to0$, the overall set on which $T_z$ is defined will have probability arbitrarily close to $P(A_z)$ and $P(A_u)$, respectively, implying that $T_z$ exists for $P_u$-almost every $u\in A_u$ and $P_Z$-almost every $z\in A_z$.
\begin{figure}[hbt]
\centering
\begin{tikzpicture}
\draw node[below] at (0,0) {$A_u^1$};
\draw node[below] at (2,0) {$A_u^2$};
\draw node[above] at (0,1.5) {$A_x^1$};
\draw node[above] at (2,1.5) {$A_x^2$};
\draw node[left] at (0,0.7) {$g(z,\cdot)$};
\draw node[right] at (2,0.7) {$g(z,\cdot)$};
\draw[->,dashed] (0,0.15) to (0,1.35);
\draw[->,dashed] (2,0.15) to (2,1.35);
\draw[<->] (0.35,1.8) to (1.7,1.8);
\draw node[above] at (1,1.8) {$T^{(1)}_z$};
\draw[->, thick] (1.8,-0.1) to (0.15,1.5);
\draw[->, thick] (0.2,-0.1) to (1.85,1.5);

\end{tikzpicture}
\caption{Underlying isomorphism structure for the ``switching procedure'' $A^1_u\mapsto A^2_x$ and $A_u^2\mapsto A_x^1$ via $T^{(1)}_z$. The bold diagonal arrows are $T^{(1)}_z\left( g(z,A^1_u)\right)$ (bottom left to top right) and $T^{(1)}_z\left( g(z,A_u^2)\right)$ (bottom right to top left).}
\label{isostructure}
\end{figure}

We now introduce the switching map from Lemma \ref{switchlemma}. Since $g(z,\cdot)$ are measure-preserving isomorphisms, it must be the case that $g(z,A^1_{u})$ and $g(z,A^2_{u})$ are disjoint and that 
\[P_{X|Z=z}(g(z,A^1_{u}))=P_{X|Z=z}(g(z,A^2_{u}))=\frac{1}{2}(\varepsilon_{u}-\delta'),\quad z\in A_z.\] For the first iteration $n=1$, let $T^{(1)}_z$ be the identity for $z\in A^2_z$; for $z\in A^1_z$ let it be such that 
\[T^{(1)}_zg(z,A^1_{u}) = g(z,A^2_{u})\qquad\text{and}\qquad T^{(1)}_zg(z,A^2_{u}) = g(z,A^1_{u}),\] i.e.~switching $g(z,A^1_{u})$ and $g(z,A^2_{u})$. Lemma \ref{switchlemma} guarantees that this is possible. The schematic for these isomorphisms is depicted in Figure \ref{isostructure}. By construction, it now holds that
\[T^{(1)}_zg(z_i,A_u^1)\neq T^{(1)}_zg(z_j,A_u^1)\qquad\text{and} \qquad T^{(1)}_zg(z_i,A_u^2)\neq T^{(1)}_zg(z_j,A_u^2)\]
for any $z_i\in A_z^1$ and $z_j\in A_z^2$. Comparing this to the definition of a one-to-one generator, we see that we need to keep this approach iteratively to make sure that each set $A_z^{\iota\wedge1}$ and $A_z^{\iota\wedge2}$ will only contain one element $z_i$, $z_j$ in the limit. This is the formal analogue of Figure \ref{Condorcet_cycle_pic}.

Now proceed iteratively for this Cantor scheme on $2^{<\mathbb{N}}$, as depicted in Figure \ref{Condorcet_cycle_pic}. At stage $n\in\mathbb{N}$ with a specific sequence $A^{\iota}_z$ for $\iota\in\{1,2\}^n$, the inductive step is to split $A^{\iota}_z$ into two closed disjoint Borel subsets of equal measure $A^{\iota\wedge 1}_z$ and $A^{\iota\wedge 2}_z$ and a residual set $N_z$ with $P_Z(N_z)=\delta^n$, in the sense that
\[P_Z(A^{\iota\wedge 1}) =P_Z(A^{\iota\wedge 2}) = \frac{1}{2}(P_Z(A^{\iota})-\delta^n)\] for the same $\delta>0$ as above. Also decompose the $2^n$-subsets of $A_u$ corresponding to all possible combinations of the complete binary tree $\{1,2\}^n$ into $2^{n+1}$ closed disjoint subsets and $2^n$ residual sets $N_u^{\iota}$ by dividing every $A_u^{\iota}$ into $A_u^{\iota\wedge 1}$ and $A_u^{\iota\wedge 2}$ of the same probability and add residual set $N_u^\iota$ of size $P_U(N_u^\iota)=(\delta')^n$ for any $\iota\in\{1,2\}^n$. 

The main idea is to again define the ``switching'' map on all of these subsets: on $A^{\iota\wedge 2}_z$ let $T^{(n)}_z$ be identical to $T^{(n-1)}_z$ and on $A^{\iota\wedge 1}_z$ let it be such that
\begin{equation}\label{generalpermutationeq}
T^{(n)}_z g(z,A^{\iota\wedge 1}_{u}) = g(z,A^{\iota\wedge 2}_{u})\quad\text{and}\quad T^{(n)}_z g(z,A^{\iota\wedge 2}_{u}) = g(z,A^{\iota\wedge 1}_{u}),
\end{equation} for every $\iota\in\{1,2\}^n$, which is possible by Lemma \ref{switchlemma}. Overall, by this construction, it holds that
\begin{equation}\label{whatwewanteq}
T^{(n)}_zg(z_i,A_u^\iota)\neq T^{(n)}_zg(z_j,A_u^\iota)
\end{equation} 
for any $z_i\in A_z^{\iota\wedge 1}$ and $z_j\in A_z^{\iota\wedge 2}$ and all $\iota\in 2^{n}, n\in\mathbb{N}$. Taking the limit $n\to\infty$ will then make the map $T_zg(z,u)$ injective in almost every $z$, i.e.~producing the required one-to-one generator, as we show below in part 1.2. Furthermore, since $g(z,\cdot)$ is by assumption a measure-preserving isomorphism and $T_z$ is invertible for all $z$ by Lemma \ref{halmoslemma}, it holds immediately that $T_zg(z\cdot)$ is one-to-one in $U$, as the composition of two injective maps is injective. We now show that this limit as $n\to\infty$ is well-defined.\\

\noindent\emph{Part 1.2: Measurability and existence of the constructed map $T_zg(z,\cdot)$}\\
We now show that the above constructed $T_zg(z,\cdot)$, defined as the limit $\lim_{n\to\infty} T_z^{(n)}g(z,\cdot)$ is measurable and well-defined for $P_Z$-almost every $z\in A_z$. The idea is to focus on $T_z$, as $g(z,\cdot)$ is always measurable and well-defined for almost every $z$ as it is a measure preserving isomorphism.
We split the argument into three parts. In the first part we show that $T_z$ is measurable; in the second part we show that it is defined almost everywhere on $A_u$ \emph{for a fixed $z\in A_z$}. After this, we show that $T_z$ exists on a set whose $P_Z$-measure can be made as large as the $P_Z$-measure of the full set $A_z$, i.e.~that $T_z$ exist for $P_Z$ almost every $z$. 

To see the measurability claim, note that the limit $T_z=\lim_{n\to\infty} T^{(n)}_z$ is measurable for fixed $z$: by Lemma \ref{switchlemma} $T_z^{(n)}$ is defined on the union of disjoint maps with Borel sets as domain and range; also on every subset $A_z^{\iota\wedge 2}$, $T_z$ is the identity. This implies that $T_z(n)$ is measurable for every $n\in\mathbb{N}$, so that $T_z$ is measurable, as it is defined as the countable limit as $n\to\infty$ \citep*[Proposition 2.7]{folland2013real}.

To see that $T_z$ is still defined almost everywhere on $A_u$, we need to show (i) that the null-sets $L^{\iota}\subset A_u^\iota$ on which $T_z^{n}$ is not defined by Lemma \ref{halmoslemma} do not accumulate and that (ii) the residual sets of size $\delta'>0$ do not accumulate if we let $\delta'\to0$. We can consider both parts independently, as the null-sets are subsets of $A_u^\iota$ and are by construction disjoint from the residual sets. Both approaches for a proof are very similar. Let us start with the null-sets.

For every $n\in\mathbb{N}$, there are $2^{n+1}$ Lebesgue null-sets $L_n^{\iota\wedge j}\subset A_u^{\iota^j}$, $j\in\{1,2\}$ on which $T^{(n)}_z$ is not defined by Lemmas \ref{halmoslemma} and \ref{switchlemma}. In particular, this implies that for any $\iota\in\{1,2\}^n$ and any $\eta>0$ $P_U(L_n^{\iota\wedge j})\leq\eta^{j+2}$. Moreover, all of the null-sets are disjoint, so that we can calculate:
\begin{align*}
\lim_{n\to\infty} P_U\left(\bigcup_{\iota\in 2^{n}} (L_n^{\iota\wedge 1}\cup L_n^{\iota\wedge2})\right)&=\lim_{n\to\infty} \sum_{\iota\in 2^n} P_U\left(L_n^{\iota\wedge 1}\right)+ P_U\left(L_n^{\iota\wedge 2}\right)\\
&= \lim_{n\to\infty}\sum_{j=0}^n 2^j\left(P_U\left(L_n^{\iota\wedge 1}\right)+ P_U\left(L_n^{\iota\wedge 2}\right)\right)\\
&\leq \lim_{n\to\infty}\sum_{j=0}^n 2^{j+1}\eta^{j+2}\\
&= \frac{2\eta^2}{1-2\eta},
\end{align*}
which tends to zero as $\eta\to0$, so that the null-sets stay a null-set in the limit. 

A similar argument holds for the residual sets: at every stage $2^{n}$ in the Cantor scheme, we partitioned \emph{every} $A_u^\iota$, $\iota\in\{1,2\}^n$ into two closed disjoint subsets $A_u^{\iota\wedge1}, A_u^{\iota\wedge 2}$ by removing a residual set $N$ of size $(\delta')^n>0$, so that we removed $n$ residual sets in total at each stage $n\in\mathbb{N}$. Therefore, at each stage $n\in\mathbb{N}$, the size of $\bigcup_{\iota\in2^n} \left(A_u^{\iota\wedge 1} \cup A_u^{\iota\wedge2}\right)$ is
\begin{align*}
P_U\left(\bigcup_{\iota\in2^n} \left(A_u^{\iota\wedge 1} \cup A_u^{\iota\wedge2}\right)\right) &= \sum_{\iota\in 2^n} P_U\left(A_u^{\iota\wedge 1} \cup A_u^{\iota\wedge2}\right)\\
&= P_U(A_u) - \sum_{j=0}^n 2^j(\delta')^{j+1}.
\end{align*}
Letting $n\to\infty$ in this Cantor scheme implies that 
\begin{align*}
\lim_{n\to\infty}P_U\left(\bigcup_{\iota\in2^n} \left(A_u^{\iota\wedge 1} \cup A_u^{\iota\wedge2}\right)\right) &=\lim_{n\to\infty} P_U(A_u) - \sum_{j=0}^n 2^j(\delta')^{j+1}\\
&=P_U(A_u) - \frac{\delta'}{1-2\delta'}.
\end{align*}
Now every $\iota\in 2^n$ is an initial segment $\tau\vert n$ for elements $\tau\in 2^{\mathbb{N}}$. Therefore, 
\[\bigcap_{\iota\in 2^n}A_u^{\iota} = \bigcap_{n\in\mathbb{N}} A_u^{\tau\vert n}\] in our Cantor scheme. This is a countable intersection of closed sets with shrinking diameter by construction, and by the completeness of the real line the limit set must contain a unique element $\{u\}\subset A_u$. 
Since $T_z$ is defined on the union of all of these single elements, this means that $T_z$ is defined on a set of measure $P_U(A_u)-\frac{\delta'}{1-2\delta'}$. Letting $\delta'\to0$ shows that every $T_z$, $z\in A_z$ can be defined on a set that has $P_U$-measure arbitrarily close to the full measure $P_U(A_u)$. The null-sets $L_n^\iota$ and the residual sets $N_u^\iota$ are disjoint for all $n\in\mathbb{N}$ and we have just shown that both are of Lebesgue measure arbitrarily close to zero, which implies that $T_z$ for fixed $z$ will be defined on every $u$ up to their union, which is a set of Lebesgue measure arbitrarily close to zero.

We now show that $T_z$ exists for all points $z$ up to a set $N$ whose $P_Z$-probability we can make arbitrarily small, exactly as the reasoning above. Recall that in the first iteration of the Cantor scheme, we partitioned $A_z$ into two closed disjoint subsets $A^1_z$ and $A_z^2$ by removing a residual set $N$ of size $\delta>0$, which is possible by Lemma \ref{splitlemma}. The argument is exactly the same as for the residual sets $N_u$, but we repeat it here. We proceeded iteratively: at every stage $n$ in the Cantor scheme, we partitioned \emph{every} $A_z^\iota$, $\iota\in\{1,2\}^n$ into two closed disjoint subsets $A_z^{\iota\wedge1}, A_z^{\iota\wedge 2}$ by removing a residual set $N$ of size $\delta^n>0$, so that we removed $n$ residual sets in total at each stage $n\in\mathbb{N}$. Therefore, at each stage $n\in\mathbb{N}$, the size of $\bigcup_{\iota\in2^n} \left(A_z^{\iota\wedge 1} \cup A_z^{\iota\wedge2}\right)$ is
\begin{align*}
P_Z\left(\bigcup_{\iota\in2^n} \left(A_z^{\iota\wedge 1} \cup A_z^{\iota\wedge2}\right)\right) &= \sum_{\iota\in 2^n} P_Z\left(A_z^{\iota\wedge 1} \cup A_z^{\iota\wedge2}\right)\\
&= P_Z(A_z) - \sum_{j=0}^n 2^j\delta^{j+1}.
\end{align*}
Letting $n\to\infty$ in this Cantor scheme implies that 
\begin{align*}
\lim_{n\to\infty}P_Z\left(\bigcup_{\iota\in2^n} \left(A_z^{\iota\wedge 1} \cup A_z^{\iota\wedge2}\right)\right) &=\lim_{n\to\infty} P_Z(A_z) - \sum_{j=0}^n 2^j\delta^{j+1}\\
&=P_Z(A_z) - \frac{\delta}{1-2\delta}.
\end{align*}
Now every $\iota\in 2^n$ corresponds to a an initial segment $\tau\vert n$ for $\tau\in 2^{\mathbb{N}}$. Therefore, 
\[\bigcap_{\iota\in 2^n}A_u^{\iota} = \bigcap_{n\in\mathbb{N}} A_u^{\tau\vert n}\] in our Cantor scheme. This is a countable intersection of closed sets with shrinking diameter by construction, and by the completeness of the real line the limit set must contain a unique element $\{z\}\subset A_z$. 
Since $T_z$ is defined on the union of all of these single elements, this means that $T_z$ is defined on a set of measure $P_U(A_u)-\frac{\delta}{1-2\delta}$. Letting $\delta\to0$ shows that $T_z$ can be defined on a set that has $P_Z$-measure arbitrarily close to the full measure $P_Z(A_z)$. This implies that $T_z$ will be defined on every $z$ up to this set of Lebesgue measure zero.
The above arguments have shown that $T_z$ is measurable, defined almost everywhere on $A_u$, and exists for $P_Z$-almost every $z\in A_z$. Since the same holds for $g(z,\cdot)$, it holds that their composition satisfies this too, which means that the map $T_zg(z,\cdot)$ is well-defined in the sense that
\[T_{z_i}g(z_i,u)\neq T_{z_j}g(z_j,u)\] for $P_U$-almost all $u\in A_u$ and $P_Z$-almost all $z_i,z_j\in A_z$ with $z_i\neq z_j$.\\ 

\noindent\emph{Part 1.3: Reduction to fixed subsets $A_z$ and $A_u$ is without loss of generality}\\
We now show that our construction can be performed for each pair of Borel sets $A_z$ and $A_u$ on which $g$ is not injective without changing the construction on other Borel sets $A'_z$ and $A'_u$. 

If the sets $A_u$ and $A_u'$ on which $g$ is not injective in $z$ are disjoint, there is nothing to prove as the maps $T_z$ are defined for each subset $A_u$ by Lemma \ref{switchlemma} and without affecting other sets $A_u'$. Furthermore, by the fact that $g(z,\cdot)$ is measure-preserving for all $z$ it must be that the images $A_x$ and $A_x'$ of $A_u$ and $A_u'$ via $g(z,\cdot)$ are disjoint.

Now consider the case where $P_U(A_u\cap A'_u)=\varepsilon>0$. This implies that the images $P_{X|Z=z}(A_x\cap A'_x)=\varepsilon>0$, by the fact that $g(z,\cdot)$ is measure preserving for all $z$. No matter if the sets $A_z$ and $A_z'$ are disjoint or not, by defining 
\[\tilde{A}_z\coloneqq A_z^1\cup A_z^2,\qquad \tilde{A}_u\coloneqq A_u^1\cup A_u^2,\qquad \tilde{A}_x\coloneqq A_x^1\cup A_x^2,\] we can apply the construction from Part 1.1 to these sets. Then the resulting construction is a one-to-one generator in both $Z$ and $U$ on all of $\tilde{A}_z$ and $\tilde{A}_u$. This holds because an injective function on a set $\tilde{A}_z$ is injective on every subset. 

This captures all possible settings and shows that the construction in Part 1.1 does not change previous constructions on other Borel sets $A_z'$ and $A_u'$. In particular, whenever Borel sets $A_u$ and $A'_u$ on which $g(z,u)$ is not injective in $Z$ are overlapping, we can simply perform the above construction on their union. The restriction of the construction to Borel (sub-) sets $A_z$ and $A_u$ is hence without loss of generality. This allows us to ignore all sets $A'_z$ and $A'_u$ for which is already a one-to-one generator in our construction, by just letting the overall map $T$ be the identity map on those sets.\\

\noindent\emph{Part 2: $P_Z$ has finitely many atoms and non-atomic parts}\\
In the case where $Z$ has a mixed distribution with $k$ many atoms, the construction from Part 1 is essentially the same except for the fact that we have a more general scheme than a Cantor scheme on $A_z$, as we now split $A_u^\iota$ at every stage into $k+2$ subsets $A_u^{\iota\wedge j}$ of the same measure, which is possible by Lemma \ref{switchlemma}. Formally, suppose there are points $z_j$ for which $P_Z(\{z_j\})>0$, $j=1,\ldots, k$, for $k\in\mathbb{N}$. We can adjust the above construction by devising a scheme on $\{1,2,\ldots, k+2\}^{<\mathbb{N}}$ as follows. 

Consider the Borel set $F\coloneqq\bigcup_{j=1}^k \{z_j\}\cup A_z$, where we denote by $A_z$ the uncountable subset of $A_z$ on which $P_Z$ is absolutely continuous by an abuse of notation. We then partition $A_z$ into two closed disjoint Borel subsets $A^{1}_z$ and $A^{2}_z$ and a residual set of measure $\delta>0$ (which we will let go to zero when proving measurability) of equal measure 
\[P_Z(A^1_z)=P_Z(A^2_z)=\frac{1}{2}(\varepsilon_Z-\delta)\] and the corresponding $A_{u}$ into $k+2$ closed disjoint Borel sets $A^{1}_{u},\ldots, A^{k+2}_{u}$ of equal measure and $k+1$ residual sets of size $(\delta')^{k+1}$, i.e.
\[P_{U}(A^1_{u}) = \ldots = P_{U}(A^{k+2}_{u})=\frac{1}{k+2}\left(\varepsilon_{u}-(\delta')^{k+1}\right).\] This is possible by Lemma \ref{splitlemma}. Then for $z_1$ let $T^{1}_{z_1}$ be the identity. For the other values $z_2,\ldots,z_k$ as well as any $z\in A^1_Z$ and $z'\in A^2_Z$ let $T^1_{z_j}$ be cyclic maps (which can be done by Lemma \ref{switchlemma}), i.e.~for $z_2$
\[T^1_{z_2}g(z_2,A^{k+2}_{u}) = g(z_2,A^1_{u}),\thickspace T^1_{z_2}g(z_2,A^{1}_{u}) = g(z_2, A^2_{u}),\thickspace\ldots,\thickspace T^1_{z_2}g(z_2, A^{k+1}_{u}) = g(z_2, A^{k+2}_{u}),\]
for $z_3$
\[T^1_{z_3}g(z_3, A^{k+1}_{u}) = g(z_3,A^1_{u}),\thickspace T^1_{z_3}g(z_3,A^{k+2}_{u}) = g(z_3,A^2_{u}),\thickspace\ldots,\thickspace T^1_{z_3}g(z_3,A^{k}_{u}) = g(z_3,A^{k+2}_{u}),\]
for $z_k$
\[T^1_{z_k}g(z_k, A^{1}_{u}) = g(z_k, A^{k}_{u}),\quad T^1_{z_k}g(z_k, A^{2}_{u}) = g(z_k, A^{k+1}_{u}),\ldots,\]
for $z\in A^1_z$
\[T^1_{z}g(z,A^{1}_{u}) = g(z,A^{k+1}_{u}),\quad T^1_{z}g(z,A^{2}_{u}) = g(z,A^{k+2}_{u}),\ldots,\]
and for $z'\in A^2_z$
\[T^1_{z'}g(z,A^{1}_{u}) = g(z,A^{k+2}_{u}),\quad T^1_{z'}g(z',A^{2}_{u}) = g(z',A^{1}_{u}),\ldots\]

Then at each iteration $n\in\mathbb{N}$ of the construction split $A^{i}_z$, $i_z\in\{1,2\}^n$, into two closed disjoint Borel subsets $A^{i_z\wedge 1}_z$ and $A^{i_z\wedge 2}_z$ of equal measure and a residual set $N$ of $P_Z$ measure $\delta>0$, i.e.\[P_Z(A^{i_z\wedge 1}_z)=P_Z(A^{i_z\wedge 2}_z)=\frac{1}{2^n}(\varepsilon_Z-\delta^n).\]  Furthermore, split $A_{u}$ into $(k+2)^{n}$ closed disjoint Borel sets $A^i_u$ of equal measure corresponding to each combination $i\in\{1,2,\ldots,k+2\}^{n}$ and the corresponding $(k+1)^n$ residual sets of size $(\delta')^{n(k+1)}$. Based on this, decompose each Borel set $A^{i}_{u}$ into $k+2$ further Borel sets $A^{i\wedge 1}_u,\ldots, A^{i\wedge k+2}_{u}$ of equal measure. This leads to $(k+2)^{n+1}$ Borel subsets of $A_u$ of equal measure, i.e.
\[P_{U}(A^{i\wedge1}_{u}) = \ldots = P_{U}(A^{i\wedge k+2}_{u})=\frac{1}{(k+2)^{n+1}}\left(\varepsilon_{u}-(\delta')^{n(k+1)}\right)\quad\forall i\in\{1,2,\ldots,k+2\}^{n}.\] Then for $z_1$ let $T^{1}_{z_1}$ be the identity. For the other values $z_2,\ldots,z_k$ as well as any $z\in A^1_z$ and $z'\in A^2_z$ let $T^1_{z_j}$ be cyclic maps (which again can be done by Lemma \ref{switchlemma}), i.e.~for $z_2$ and every $i\in\{1,2,\ldots,k+2\}^{n}$
\[T^{n}_{z_2}g(z_2,A^{i\wedge k+2}_{u}) = g(z_2,A^{i\wedge1}_{u}),\quad\ldots,\quad T^{n}_{z_2}g(z_2,A^{i\wedge k+1}_{u}) = g(z_2,A^{i\wedge k+2}_{u}),\]
for $z_3$
\[T^{n}_{z_3}g(z_3,A^{i\wedge k+1}_{u}) = g(z_2,A^{i\wedge1}_{u}),\quad\ldots,\quad T^{n}_{z_2}g(z_2, A^{i\wedge k}_{u}) = g(z_2,A^{i\wedge k+2}_{u}),\]
for $z_k$
\[T^{n}_{z_k}g(z_k,A^{i\wedge 1}_{u}) = g(z_k,A^{i\wedge k}_{u}),\quad T^{n}_{z_k}g(z_k,A^{i\wedge 2}_{u}) = g(z_k,A^{i\wedge k+1}_{u}),\ldots,\]
for $z\in A^{i_z\wedge1}_z$
\[T^{n}_{z}g(z,A^{i\wedge 1}_{u}) = g(z,A^{i\wedge k+1}_{u}),\quad T^{n}_{z}g(z,A^{i\wedge 2}_{u}) = g(z,A^{i\wedge k+2}_{u}),\ldots,\]
and for $z'\in A^{i_z\wedge2}_z$
\[T^{n}_{z'}g(z',A^{i\wedge 1}_{u}) = g(z',A^{i\wedge k+2}_{u}),\quad T^{n}_{z'}g(z',A^{i\wedge 2}_{u}) = g(z',A^{i\wedge1}_{u}),\ldots\]
Then again as before the limit as $n\to\infty$ will yield a well-defined $T_z$ in the sense that it is measurable and is defined for $P_U$-almost every $u\in A_u$ and $P_Z$-almost every $z\in A_z$. The proof of this is exactly the same as in part 1.2 if we denote $\delta''\coloneqq (\delta')^{k+1}$ and is hence omitted. \\

\noindent\emph{Part 3: Countably many atoms and purely atomic $P_Z$}\\
From Part 2, one can obtain the result for a countably infinite number of atoms and an absolutely continuous part by letting $k\to\infty$ and devising a classical Lusin scheme $\mathbb{N}^{<\mathbb{N}}$ instead of the more restrictive scheme $\{1,2,\ldots, k+2\}^{<\mathbb{N}}$. Note that the space $\{1,2,\ldots,k+2\}^{\mathbb{N}}$ approaches the Baire space $\mathbb{N}^{\mathbb{N}}$ as $k\to\infty$. Both spaces are of the same cardinality \citep*[Theorem 7.8]{kechris1995classical}. The following paragraphs contain the details.

The same measurability result for $T_z$ at a fixed $z$ holds as above: $T_z=\lim_{n\to\infty} T_z^{(n)}$ is a countable limit, and each $T_z^{(n)}$ is measurable because it is defined on a countable union of Borel sets mapping to a countable union of Borel sets.

To show the fact that $T_z$ is defined almost everywhere on $A_u$, recall that in part 2 we had a union of $k+2$ of null-sets at every stage $n$ of the form $\bigcup_{n\in\mathbb{N}}\bigcup_{\iota\in \{1,\ldots,k+2\}^{n}} \bigcup_{j=1}^{k+2}N^{\iota\wedge j}$ and a union of $k+1$ residual sets. Let us consider the null-sets $L_n^\iota$ first. Since each $L_n^\iota$ is a Lebesgue null-set, it holds for every $\iota\in \{1,\ldots,k\}^{n}$ and $\eta>0$ that $P_U(L_n^{\iota\wedge j})\leq k^{-2j-k}$.

 The same reasoning holds now, because we let $k\to\infty$ at every stage $\mathbb{N}^n$. Since all $L_n^{\iota\wedge j}$ are disjoint, we have
 \begin{align*}
\lim_{k\to\infty}\lim_{n\to\infty} P_U\left(\bigcup_{\iota\in \{1,\ldots,k\}^{n}} \bigcup_{j=1}^kL_n^{\iota\wedge j}\right) &= \lim_{k\to\infty}\lim_{n\to\infty}\sum_{\iota\in \{1,\ldots,k\}^n}\sum_{j=1}^k P_U(L_n^{\iota\wedge j})\\
 &= \lim_{k\to\infty}\lim_{n\to\infty}\sum_{i=0}^n k^i \sum_{j=1}^k P_U(L_n^{\iota\wedge j})\\
 &\leq \lim_{k\to\infty}\lim_{n\to\infty}\sum_{i=0}^n k^{i+1} k^{-2i-k}\\
 &=\lim_{k\to\infty}\sum_{i=0}^\infty k^{-i-k+1}\\
 &= \lim_{k\to\infty}\frac{1}{(k-1)k^{k-2}}\\
 &=0.
 \end{align*}
This implies that the null-sets do not accumulate. 

The residual sets $N_u^\iota$ do not accumulate either, as at every stage $n$ their size is $(\delta')^{n(k+1)}$ and we are taking the limit $k\to\infty$, which means that the residual sets become a null-set at every stage $n$. We can therefore take their union with the null-sets caused by Lemma \ref{halmoslemma} to conclude that both will be a null-set in the limit. Since we now have a Lusin scheme of closed disjoint sets $A_u^{\iota}$ $\iota\in\mathbb{N}$ at every stage $n$, it follows again that the countable sequence 
\[\lim_{n\to\infty}\bigcap_{\iota\in\mathbb{N}^n} A^{\iota}\equiv \bigcap_{n\in\mathbb{N}} A^{\tau\vert n}\] converges to a single point for $\tau\in \mathbb{N}^{\mathbb{N}}$. Since $T_z$ is defined on all of these points, it follows that it is defined for $P_U$-almost every $u\in A_u$. 

What is left is to show is existence of $T_z$ for $P_Z$-almost every $z$. But this follows from exactly the same reasoning as part 2. In particular, $T_z$ is defined on every atom. A potential set $A_z$ on which $P_Z$ is non-atomic will be dealt with in the exact same way as in part 1.2 by defining one residual set in every stage of $P_Z$-measure $\delta>0$ and letting $\delta\to0$. This shows that $T_z$ is also well-defined in this last case in the sense that
\[T_{z_i}g(z_i,u)\neq T_{z_j}g(z_j,u)\] for $P_U$-almost all $u\in A_u$ and $P_Z$-almost all $z_i,z_j\in \bigcup_{k=1}^\infty \{z_k\}\cup A_z$ with $z_i\neq z_j$.\\

In all cases we have thus constructed the sought-after function by letting $T_z$ be the identity map on all other sets except those sets $A_z$ on which $g$ is not one-to-one in $z$. Part 1.3 shows that this does not affect the construction on other Borel sets on which $g$ is not one-to-one in $Z$.
We have therefore provided a construction that makes $g(z,u)$ a one-to-one generator in the case where $P_{X|Z=z}$ is non-atomic for $P_Z$-almost all $z$. Moreover, as mentioned above, by the fact that we can choose $g$ to be injective in $U$ for all $z$ and that $P_Z$-almost all $T_z$ are invertible, this construction immediately implies that $T_zg(z\cdot)$ is invertible in $U$ for $P_Z$-almost all $z$. This construction is measurable and well-defined in all cases as shown above, so that we can apply Lemma \ref{pearlslemma} to finish the proof.
\end{proof}

\bibliography{main_bib}

\end{document}